\documentclass[%
 aip,
 amsmath,amssymb,
 reprint,%
]{revtex4-1}

\usepackage{graphicx}
\usepackage{dcolumn}
\usepackage{bm}
\usepackage{xcolor}

\usepackage[utf8]{inputenc}
\usepackage[T1]{fontenc}
\usepackage{mathptmx}
\usepackage{etoolbox}

\makeatletter
\def\@email#1#2{%
 \endgroup
 \patchcmd{\titleblock@produce}
  {\frontmatter@RRAPformat}
  {\frontmatter@RRAPformat{\produce@RRAP{*#1\href{mailto:#2}{#2}}}\frontmatter@RRAPformat}
  {}{}
}%
\makeatother

\begin{document}


\title{Thermometry of an optically levitated nanodiamond}

\author{François Rivière}
\thanks{equally contributed authors}
\author{Timothée de Guillebon}
\thanks{equally contributed authors}
\author{Léo Maumet}
\affiliation{Universit\'e Paris-Saclay, CNRS, ENS Paris-Saclay, CentraleSup\'elec, LuMIn, 91405, Orsay, France}
\author{Gabriel Hétet}
\affiliation{Laboratoire De Physique de l’\'Ecole Normale Supérieure, \'Ecole Normale Sup\'erieure, PSL Research University, CNRS, Sorbonne Universit\'e, Universit\'e de Paris , 24 rue Lhomond, 75231 Paris Cedex 05, France}
\author{Martin Schmidt}
\author{Jean-Sébastien Lauret}
\author{Loïc Rondin}
\affiliation{Universit\'e Paris-Saclay, CNRS, ENS Paris-Saclay, CentraleSup\'elec, LuMIn, 91405, Orsay, France}
\email{loic.rondin@universite-paris-saclay.fr}


\date{\today}

\begin{abstract}
Using the spin properties of nitrogen-vacancy (NV) centers in levitated diamond, we characterize the absorption of single nanodiamonds. We first calibrate the thermometry response of the NV centers embedded in our nanodiamonds. Then, using this calibration, we estimate the absorption cross-section of single levitated nanodiamonds. We show that this absorption is extrinsic and dominated by volumic effects. 
Our work opens the way to diamond materials optimization for levitation quantum experiments. It also demonstrates optical levitation as a unique platform to characterize material thermal properties at the nanoparticle level.
\end{abstract}

\maketitle

\section{Introduction}%
Levitated diamonds in vacuum have been proposed as a promising platform for spin-mechanics experiments~\cite{Perdriat2021M}. Indeed, doping diamonds with NV colors centers, a spin-active defect of the diamond lattice, provides a  controllable internal quantum degree of freedom. This control can then be used for spin-cooling of the diamond motion~\cite{Delord2020N}, matter-waves experiments~\cite{Bose2017PRL,Wood2022PRA} and non-classical state preparations. 
However, a critical limitation to the overall development of this approach is the severe heating reported for levitated diamonds~\cite{Delord2017APL,Hoang2016NC,Pettit2017JOSABJ,OBrien2019APL}, an effect particularly significant for optical levitation experiments. Understanding the mechanisms related to laser absorptions is thus of prime importance. 

In this context, we report on the thermometry of optically levitated nanodiamonds. Our study lies on the temperature dependence of the NV center spin resonance. Thus, we first calibrate the temperature response of the NV centers embedded in our particular nanodiamonds. We then use this calibration to measure the heating coefficient of optically levitated nanodiamonds and study how the diamond size impacts it. Finally, we assess our nanodiamonds' absorption cross-section and discuss the potential mechanisms responsible for this absorption.
Our work demonstrates the power of optical levitation to study material properties, such as absorption, at the single nanoparticle level. It also constitutes an essential step to optimize diamond material for quantum spin-levitation experiments. 
\begin{figure}[htb]
        \centering
        \includegraphics[width=0.99\linewidth]{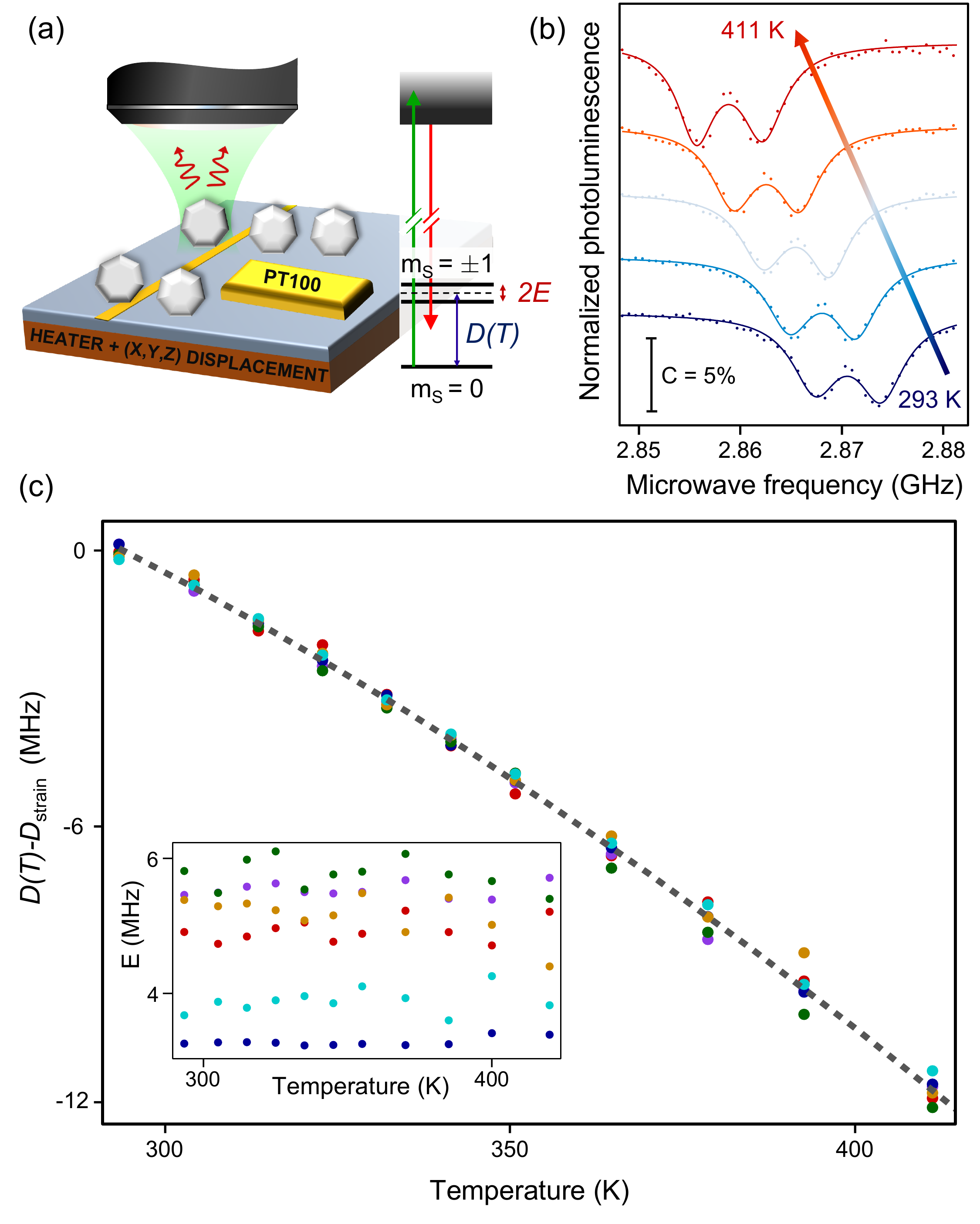}
        \caption{Caracterization of NV thermometry. (a) Experimental apparatus used for the realization of ESR under heating. The schematized NV center electronic level system is presented. (b) ESR experiments realized at five different temperatures for a nanodiamond. We observe a shift of the center $D$ of the two ESR lines, while the splitting $E$ between them remains constant. The different curves are shifted for clarity. The black bar indicates 5~\% contrast. Typical observed ESR contrast is 5 to 10~\%. (c) Evolution of $D(T)$ corrected from $D_\text{strain}$ in function of the temperature for 6 different nanodiamonds. The dotted line correspond to the polynomial equation from Toyli et al~\cite{Toyli2012}. The inset shows the stability of the parameter $E$ under heating for each nanodiamond.}%
        \label{fig1}
\end{figure}

\section{NV thermometry of nanodiamonds}%
\label{sec:nv_thermometry_of_nanodiamonds}

The NV color center in diamond corresponds to an optically addressable spin $S=1$ system in which the $|m_s=0\rangle$ and $|m_s=\pm 1 \rangle$ sublevels are split due to a spin-spin coupling along with a splitting $2E$ of the $|m_s=\pm 1 \rangle$ sublevels, particularly noticeable for nanodiamonds due to a strong internal strain. A simplified energy levels diagram of the NV center is presented in Figure~\ref{fig1}-(a).
Under green laser excitation, the NV center emits red photoluminescence which is spin-dependent, allowing the optical detection of the electron spin resonance (ESR). 
From a bi-Lorentzian fit, we determine the value of the zero field splitting $D$ between the $|m_s=0\rangle$ and $|m_s=\pm 1 \rangle$ sublevels. This $D$ parameter is temperature dependent due to electron-phonon coupling and thermal expansion of the diamond lattice~\cite{Doherty2014,Ivady2014}. Its temperature evolution $D(T)$ has been thoroughly studied, especially for bulk diamond, and empirical laws have been proposed~\cite{Doherty2014,Ivady2014,Toyli2012,Acosta2010,Plakhotnik2014}. In this context, we aim at calibrating the temperature of our specific batch of nanodiamonds. This is particularly relevant for nanodiamonds due to a strong dispersion of $D$ at room temperature between different particles, linked to changes in strain, surface quality or geometry~\cite{Foy2020,Plakhotnik2014}.


To gain knowledge on the behavior of our heavily doped nanodiamonds (FND-br100 diamonds, FND Bitotech) under heating, the evolution of $D(T)$ with temperature is studied. To this end, the nanodiamonds are spin-coated on a quartz coverslip which is placed on a 3-axis displacement stack to perform scanning. Using a green laser excitation (Laser Quantum GEM, $\lambda=532$~nm), the red photoluminescence of the NV center is collected as depicted in Figure~\ref{fig1}-(a). The microwave signal needed to realize ESR measurement is brought by a copper wire in the vicinity of the nanodiamonds. A heater is placed underneath the sample and the temperature is controlled thanks to a PID device (Lakeshore Instruments) coupled to a Pt100 thermal sensor glued to the sample with thermal paste. A temperature correction is applied to take into account heat losses (see Supplementary Information).

For a given nanodiamond, we observe a regular shift of the ESR lines under heating from ambiant temperature to 411~K, as presented on Figure~\ref{fig1}-(b). To confirm this behavior, we performed the same measurement for 6 different nanodiamonds. Fitting the ESR spectrum, we obtain the evolution of $D(T)$ for these nanodiamonds. To take into account the intrinsic strain $D_\text{strain}$ of each nanodiamond, we fit the data with the polynomial equation previously determined by Toyli et al~\cite{Toyli2012} (see Supplementary Information). We then subtract $D_\text{strain}$ to obtain the data presented on Figure~\ref{fig1}-(c).

From these results, we get few interesting conclusions. First of all, we confirm the previous observations related to the evolution of $D$ with temperature. In particular, we observe a very good agreement with the polynomial temperature dependance determined previously by Toyli et al~\cite{Toyli2012}. As a consequence, we use this polynomial determination, corrected from the offset $D_\text{strain}$, to obtain the temperature of the levitated nanodiamond in the following. Besides, we observe a small dispersion at each temperature. By calculating the standard deviation for $D$ for each point and by dividing it by the derivative of the polynom, we determine the mean uncertainty in the determination of the temperature to be $\Delta T=2.0$~K. In addition, we observe that the value of the strain parameter $E$ is stable under heating, as presented on the inset of Figure~\ref{fig1}-(c).

\section{Internal temperature of a single levitated nanodiamond}%
\begin{figure}[htb]
        \centering
        \includegraphics[width=0.8\linewidth]{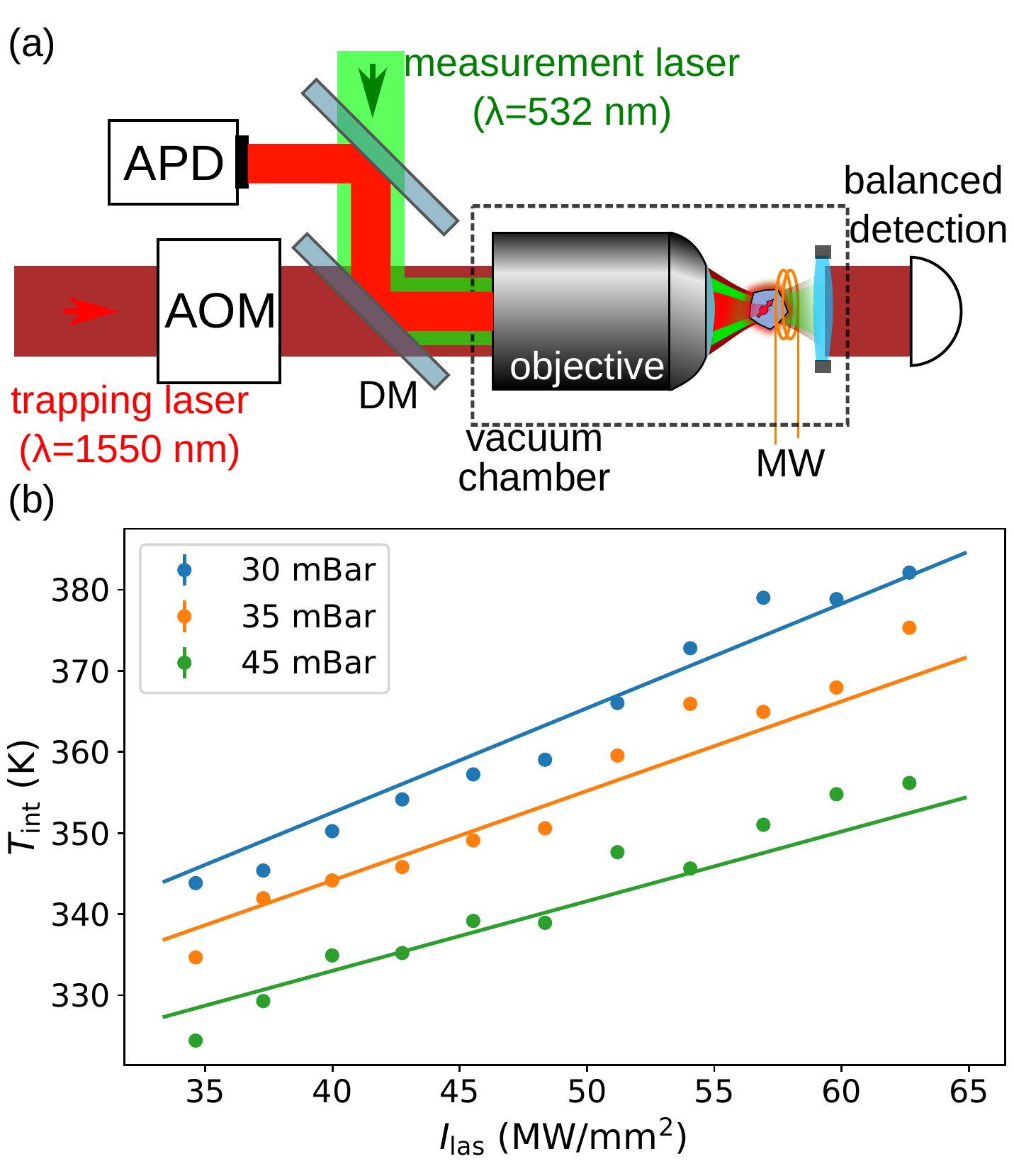}
        \caption{Heating of optically levitated nanodiamonds. (a) Experimental setup used for the measurement of internal temperature of the levitated nanodiamonds. DM stands for \textit{Dichroic Mirror} and MW for \textit{Microwave}. (b) Typical evolution a nanodiamond internal temperature recorded as a function of the laser power density, for different gas pressures. The plain lines correspond to fits using equation~(\ref{eq:Tint}).}%
        \label{fig2}
\end{figure}

\label{sec:internal_temperature_of_a_single_levitated_nanodiamond}
Thus, we can use the temperature dependence of the NV spin resonance to study the internal temperature of levitated nanodiamonds. To that purpose, we use the experimental setup depicted in Figure~\ref{fig2}-(a). A single nanodiamond is trapped at the focus of a high-NA microscope objective using a high-power infrared laser (Keopsys CEFL-KILO, $\lambda$=1550~nm). An acousto-optics modulator (AOM) allows to finely control the trapping laser intensity. The photoluminescence of the NV centers hosted by the levitated nanodiamond are addressed with the green laser, copropagating with the infrared laser. The luminescence is collected through the trapping objective and directed onto an avalanche photodiode (APD). The spin resonance of the NV centers is excited with a copper antenna rolled around the optical trap. The system is enclosed inside a vacuum chamber, and the residual gas pressure $p_\text{gas}$ inside the chamber is measured using a capacitance gauge (Pfeiffer vacuum, CMR 361). Finally, the trapping laser also serves to measure the particle dynamics, using a balanced detection after the vacuum chamber~\cite{Gieseler2012PRL,Riviere2022AP}.
Note, that in the present study, we keep the green laser at a low power such that we only study the absorption by the infrared trapping laser. 

The internal temperature of the levitated nanodiamond is determined from the analysis of the electron spin resonance spectra, as discussed in the first section. An example of the measured internal temperature of a levitated nanodiamond as a function of the trapping infrared laser power for different gas pressures is shown in Figure~\ref{fig2}-(b). We determine the value of $D_\text{strain}$ from the polynomial equation of $D(T)$ by extrapolating the value of the diamond internal temperature at vanishing laser power.
We note that typically we measure temperatures up to 650~K before loosing the particle from the trap. Also, independently of their internal temperature, nanodiamonds always escape from the trap at pressure between 1 and 10~mbar. Thus, in the following we always keep the nanodiamonds in their stability region, \textit{i.e} controlling the laser power and gas pressure to keep their temperatures below $\simeq 550$~K, and the gas pressure above 15~mbar.


As shown in Figure 2-(b), we observe a linear increase of the internal temperature with the laser power density $I_\text{las}$, and a heating related to a decrease in residual gas pressure $p_\text{gas}$.
Indeed, the internal temperature of levitated particles originates from a competition between the different thermal processes involved. In the pressure regime explored in our study, thermal black-body radiations can be safely neglected over the gas conduction processes~\cite{Chang2010,Frangeskou2018NJP}. Thus, the final particle temperature corresponds to the balance between the absorption power $\mathcal P_\text{abs}$ and the gas conduction $\mathcal P_\text{cond}$. 
To describe the absorption by the particle, we introduce the absorption cross-section $\sigma_\text{abs}$, such that 
\begin{equation}
        \mathcal P_\text{abs} = \sigma_\text{abs}(\lambda)I_\text{las}\, .
\end{equation}
The absorption mechanism may be related to the purity of the diamond matrix, and is impacted by the presence of internal crystallographic defects. For instance, the diamond particles we use are rich in nitrogen impurities ([N]$\sim$200~ppm) and NV centers ([NV]$\sim$2~ppm) and may contain vacancies and structural defects. All these defects may contribute to light absorption with a strength depending on the considered wavelength. In this picture, we expect the absorption cross-section to be proportional to the particle volume $V_\text{part}$.
Alternatively, the absorption may also arise from the surface quality where the presence of graphite would significantly contribute to laser heating. If so, absorption cross-section would scale with the particle surface $S_\text{part}$.

Concerning the thermal power dissipated by gas conduction in the free-molecular regime, a condition that is verified for the investigated pressures, it writes~\cite{Liu2006APB}
\begin{equation}
        \mathcal P_\text{cond} = \alpha_\text{acc} S_\text{part} \dfrac{p_\text{gas}\bar c}{8T_0}\dfrac{\gamma+1}{\gamma -1}\left(T_\text{int}-T_0\right)\, ,
\end{equation}
where $T_0=294$~K  and $T_\text{int}$ are the ambient and particle internal temperature, $\bar c\approx 503$~m/s is the mean thermal speed of the gas molecules and $\gamma=7/5$ is the gas specific ratio.
$\alpha_\text{acc}$ is the particle accommodation coefficient. In a recent work~\cite{Riviere2022AP}, we showed that for nanodiamonds $\alpha_\text{acc}\approx 1$. Thus, in the following, we assume that $\alpha_\text{acc}=1$ for all particles. 

\begin{figure}[htb]
        \centering
        \includegraphics[width=.95\linewidth]{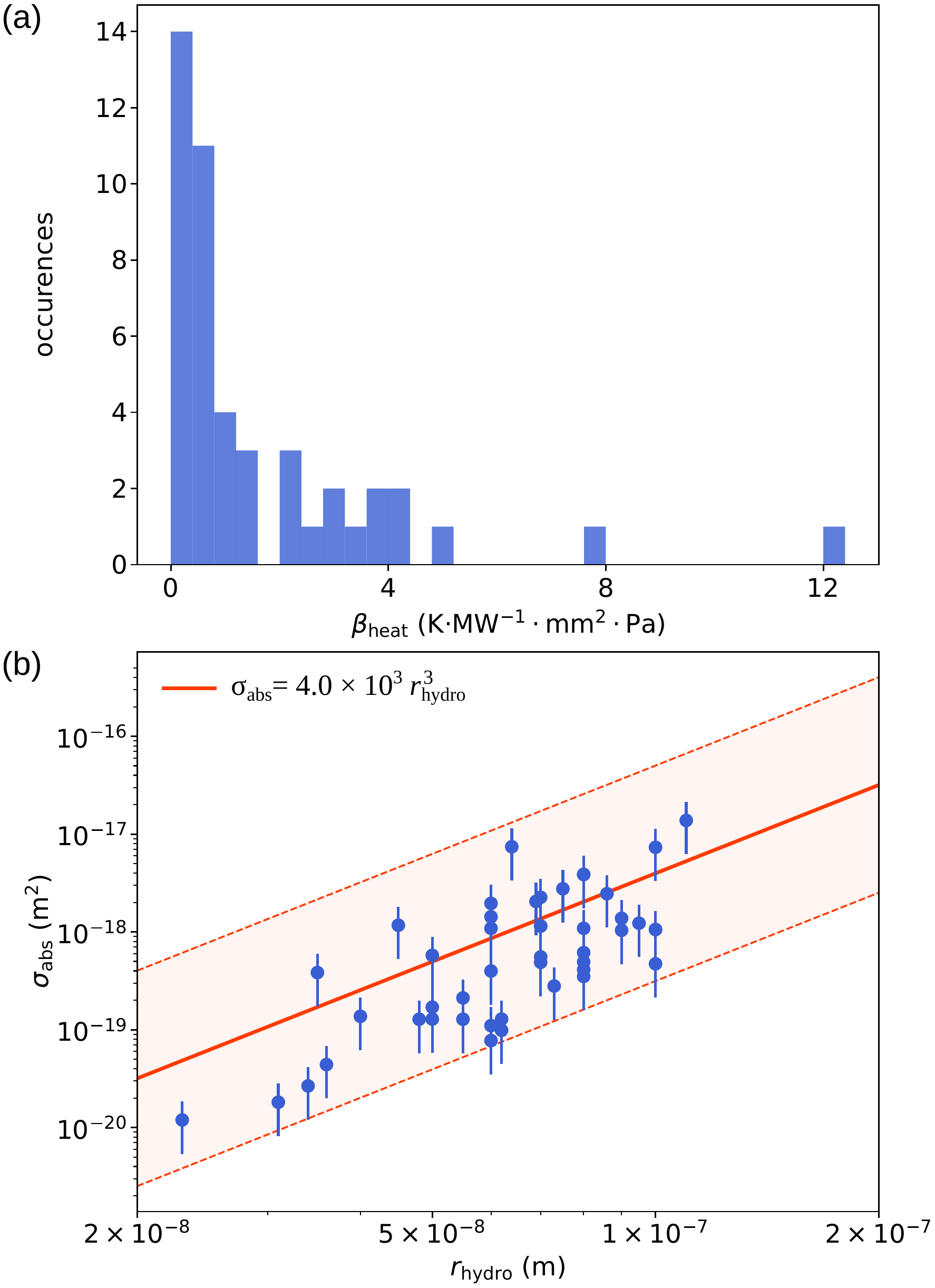}
        \caption{(a) Histograms of the heating coefficient $\beta_\text{heat}$ recorded on a batch of 46 nanodiamonds. (b) Absorption cross-section estimated from the nanodiamonds batch presented in (a) as a function of the nanodiamond hydrodynamic radius. The orange plain line correspond to the best fit of a $r_\text{hydro}^3$ function. Orange shadowed area corresponds to $r_\text{hydro}^3$ functions delimiting a confidence interval at 95.5~\%.}%
        \label{fig3}
\end{figure}

At thermal equilibrium, $\mathcal P_\text{abs}=\mathcal P_\text{cond}$, such that the particle temperature writes: 
\begin{equation}
        T_\text{int} = T_0 + \beta_\text{heat}\dfrac{I_\text{las}}{p_\text{gas}}\, . 
        \label{eq:Tint}
\end{equation}
where
\begin{equation}
        \beta_\text{heat}= \dfrac{\sigma_\text{abs}}{S_\text{part}}\dfrac{2T_0}{\bar c}\dfrac{\gamma-1}{\gamma+1}\, ,
        \label{eq:beta}
\end{equation}
characterizes the heating coefficient of a given nanodiamond. We can then determine this heating coefficient by a fit of the data presented in the figure~\ref{fig2}-(b). We note that in the figure, the fit is done simultaneously on the pressure and on the laser power.
To get further insights into the heating mechanism in nanodiamonds, we measured the heating coefficient $\beta_\text{heat}$ on 46 nanodiamonds. The distribution of this  $\beta_\text{heat}$ is shown in Figure~\ref{fig3}-(a). In a simplistic model, we expect that if the heating coefficients were dominated by a uniform surface absorption, for example  by a surface graphite layer, since $\beta_\text{heat}\propto\dfrac{\sigma_\text{abs}}{S_\text{part}}$ (see eq.(\ref{eq:beta})), then they would be size independent. This is in contrast with the wide dispersion we observe here. Thus, at first order, infrared laser absorption can not be attributed to a generic diamond surface. Nevertheless, we can not rule, at this point, the presence of localized absorption defects on the surface, such as graphite patches.

To sharpen our analysis of the absorption mechanisms in nanodiamonds, we propose to estimate individual diamonds' absorption cross-section. 
We note that the only unknown parameter in equation (\ref{eq:beta}), linking $\beta_\text{heat}$ to $\sigma_\text{abs}$, is the particle surface $S_\text{part}$. 
If this quantity is particularly challenging to address, one can first estimate it by assuming that the nanodiamond's characteristic size is its hydrodynamic radius $r_\text{hydro}$, such that  $S_\text{part}\approx 4 \pi r_\text{hydro}^2$. The hydrodynamic radius of a particle is the effective radius that would have a spherical particle with the same damping coefficient. While the hydrodynamic radius provides only a rough estimation of the nanodiamond size, it presents the advantage of being easy to measure. Indeed, by recording the nanodiamond dynamics and computing the associated power spectral density, we retrieve the damping coefficient $\Gamma$ of the nanodiamond. The hydrodynamic radius is then directly proportional to $1/\Gamma$~\cite{Hoang2016PRL,Riviere2022AP} (see Supplementary Information). 

By measuring the hydrodynamic radius for the batch of nanodiamonds presented in Figure~\ref{fig3}-(a), we compute the associated absorption cross-section 

\begin{equation}
       \sigma_\text{abs}=\beta_\text{heat}r_\text{hydro}^2 \dfrac{\pi\bar c}{2T_0}\dfrac{\gamma+1}{\gamma-1}\, .
        \label{eq:sigmaabs}
\end{equation}
 
As shown in Figure~\ref{fig3}-(b), we observe that, as expected, the absorption cross-section increase with the particle size, despite an important dispersion. 
To highlight the leading absorption mechanism, we try to fit the data with a function of the form $a_h r_\text{hydro}^n$. When $a_h$ and $n$ are used as free parameters, we find $n\approx 3.3$. Also, the fit is better when we enforce $n=3$ than when $n=2$ (see Supplementary Information). This result supports that volume absorption is dominant in heating nanodiamonds under NIR illumination. We then attribute the broad dispersion of the data to the fact that the nanodiamonds are not perfectly homogeneous and that local defects may impact the diamond absorption.

From the best fit with the function $a_h r_\text{hydro}^3$ (orange line in figure~\ref{fig3}-(b)), we find $a_h=4.0\times10^3$~m$^{-1}$. To take the dispersion of the data into account, we calculate the standard deviation between the data and the best fit. We thus obtain a confidence interval at 95.5 \% framing our data (shadowed area in Figure~\ref{fig3}-b), corresponding to the functions $a_\pm r_\text{hydro}^3$ with $a_-=3.1\times10^{2}$~m$^{-1}$ and $a_+=5.0\times10^{4}$~m$^{-1}$.

Since the absorption is volumic, assuming the particle is homogeneous, we can estimate the imaginary part of the diamond's relative dielectric constant $\epsilon=\epsilon'+i\epsilon''$. Indeed, for a uniform particle in the Rayleigh regime ($r_\text{part}\ll \lambda$) the absorption cross-section writes
\begin{equation}
        \sigma_\text{abs}=6\pi\dfrac{V_\text{part}}{\lambda}\Im\left(\dfrac{\varepsilon-1}{\varepsilon+2}\right)\, .
\end{equation}
For a real part of the diamond relative dielectric constant $\epsilon'=5.7$, we get $\epsilon''$ between $2.4\times 10^{-4}$ and $3.9 \times 10^{-2}$ from the values of $a_\pm$ obtained previously. Computing this dielectric constant is then important to estimate the bulk absorption, which is the physical quantity measured at the macroscopic scale on bulk material. We thus estimate a bulk absorption coefficient ranging from 4 to 662~cm$^{-1}$ for the studied nanodiamonds. 

Interestingly, we note that while broad ranges of absorption have been reported for different diamonds~\cite{Walker1979RPP,Zaitsev2001}, the ultimate intrinsic absorption cross-section of diamond at the considered wavelength is below $0.01$~cm$^{-1}$~[\onlinecite{Webster2015JOSABJb}], showing room for improvement. Also, we highlight that our measured absorption cross-section at $\lambda=1550$~nm is very similar to the absorption cross-section $\sigma_\text{abs}^\text{NV}\approx 3\times 10^{-18}$~m$^{-2}$ reported in similar diamonds due to a green 532~nm laser absorption by NV centers~\cite{Fujiwara2021SA,Wee2007JPCA}. 
Since we use a weak green excitation laser power density, this confirms that the infrared absorption is dominant in the heating mechanism of our levitated nanodiamonds.

\section{Discussion on nanodiamonds absorption}
We have shown that volume effects dominate diamond absorption at $\lambda=1550$~nm. It is thus interesting to discuss the nature of this absorption. 
While the observed heating coefficients for optically levitated diamonds are in line with previous reports~\cite{Hoang2016NC,Pettit2017JOSABJ,Riviere2022AP}, it is, as discussed previously, orders of magnitude above the expected absorption value for pure bulk diamond crystals. Thus, the observed absorption is extrinsic and comes from defects of the nanodiamond lattice. 
Interestingly, there are few reports for defects with absorption around $\lambda=1550$~nm~\cite{Zaitsev2001}.
The main impurities of our nanodiamonds, nitrogen atoms and  NV centers, are expected to have a weak absorption at this wavelength. Indeed, single nitrogen atoms have both an absorption band in the UV, with an absorption tail that decreases strongly before reaching the NIR region, and vibronic structures in the infrared above 2~$\mu$m. The absorption of the NV centers is reported to be mostly in the visible or in the infrared near $\lambda=1042$~nm. 
Nitrogen aggregates may present an absorption band near the considered wavelength but they are believed to be unlikely at our nanodiamond's nitrogen content. Alternatively, defects of the diamond lattice, grain boundaries or dislocations, may produce a broadband absorption. Nevertheless, nanodiamonds made from milling are generally mostly monocrystalline. Finally, indirect absorption related to the NV centers, due to the complex photophysics of the NV centers illuminated with infrared light~\cite{Ji2016} may also be a source of absorption. 

To get deeper insights on this question, it would be interesting to extend the present study to different absorption wavelengths~\cite{Pettit2017JOSABJ,Delord2017APL} and to other diamond materials. 
Of specific interest would be to characterize the absorption as a function of the nitrogen or NV contents and test different diamond matrices such as CVD-grown nanodiamonds~\cite{Feudis2020AMI}.  Ultimately, a fine characterization of the absorption in ultra-pure nanodiamonds~\cite{Frangeskou2018NJP} doped with a single NV center would be an important step to optimize diamond material for quantum spin-levitodynamics experiments. Such studies would also provide valuable information on NV centers' and diamonds' photophysics and the coupling to lattice vibration.

\section{Conclusion}%
\label{sec:conclusion}
Finally, we have characterized the thermal response of NV centers in nanodiamonds toward their use as sensitive thermometer for levitated diamonds. Using this single spin thermometer, we have measured the heating coefficient of levitated nanodiamonds under heating due to the infrared trapping laser power. 
Assuming the nanodiamonds as spheres, we estimate their absorption cross-section at the single-particle level. We highlight that the absorption is proportional to the particle volume from measurements on more than forty nanodiamonds. Also, we show an absorption orders of magnitude larger than the one expected for pure diamond lattice, highlighting the extrinsic nature of the absorption. This result is significant for diamond material optimization toward using levitated nanodiamonds in hybrid spin-levitodynamics experiments. Extending our work to diamonds from different fabrication processes, sizes, and impurities contents will provide critical insights into the determination of diamond absorption causes. 

Also, our work shows that the optical levitation of particles is a powerful tool for characterizing the thermal properties of materials at the nanoparticle level. Our approach, coupled to a fine estimation of the particle size~\cite{Ricci2019NL}, may enable precise determination of the absorption cross-section of single nanoparticles. Furthermore, in the absence of NV centers, it could be naturally extended by measuring the particle temperature from its dynamics in the optical trap~\cite{Millen2014,Riviere2022AP}, making our approach available for any material.

\begin{acknowledgments}

This work is supported by the Investissements d’Avenir of LabEx PALM (ANR-10-LABX-0039-PALM), by the Paris \^ile-de-France Region in the framework of DIM SIRTEQ, and by the ANR OPLA project (ANR-20-CE30-0014).
We thank Jean-François Roch for discussions. 
\end{acknowledgments}

\section*{Data Availability Statement}

The data that support the findings of this study are available from the corresponding author upon reasonable request.

\section*{Conflict of interest}
The authors have no conflicts to disclose.

\bibliography{biblio}

\end{document}



\title{Supplementary Information for Thermometry of an optically levitated nanodiamond}

\author{François Rivière}
\thanks{equally contributed authors}
\author{Timothée de Guillebon}
\thanks{equally contributed authors}
\author{Léo Maumet}
\affiliation{Universit\'e Paris-Saclay, CNRS, ENS Paris-Saclay, CentraleSup\'elec, LuMIn, 91405, Orsay, France}
\author{Gabriel Hétet}
\affiliation{Laboratoire De Physique de l’\'Ecole Normale Supérieure, \'Ecole Normale Sup\'erieure, PSL Research University, CNRS, Sorbonne Universit\'e, Universit\'e de Paris , 24 rue Lhomond, 75231 Paris Cedex 05, France}
\author{Martin Schmidt}
\author{Jean-Sébastien Lauret}
\author{Loïc Rondin}
\affiliation{Universit\'e Paris-Saclay, CNRS, ENS Paris-Saclay, CentraleSup\'elec, LuMIn, 91405, Orsay, France}
\email{loic.rondin@universite-paris-saclay.fr}


\date{\today}

\maketitle

\section{Model for the temperature dependance of the zero-field splitting $D$}

In the experiment displayed in the Figure 1 of the main text, we model the evolution of $D$ with temperature using the polynom proposed by Toyli et al~\cite{Toyli2012}. To take into account the heat losses at the level of the nanodiamond on the sample, we correct the temperature with a coefficient $\alpha$ such that :

\begin{equation}
D(T)=a_0+a_1(\alpha T)+a_2(\alpha T)^2+a_3(\alpha T)^3\, ,
\end{equation}
where $a_0$ and $\alpha$ are free parameters and $a_1=9.7\times10^{-5}$ GHz/K, $a_2=-3.7\times10^{-7}$ GHz/K$^2$ and  $a_3=1.7\times10^{-10}$ GHz/K$^3$. The values of $\alpha$ and $a_0$ extracted from the fit allow us to respectively correct the effective temperature and obtain $D_\text{strain}$ for each nanodiamond.

For the nanodiamonds studied in the optical trap in the rest of the study, the measurement of $D$ allows us to extract the temperature thanks to the polynom of Toyli et al~\cite{Toyli2012} 
\begin{equation}
D(T)=a_0+a_1T+a_2T^2+a_3T^3\, ,
\end{equation}
with $a_0=2.8697$ GHz, $a_1=9.7\times10^{-5}$ GHz/K, $a_2=-3.7\times10^{-7}$ GHz/K$^2$ and  $a_3=1.7\times10^{-10}$ GHz/K$^3$. The value of $D_\text{strain}$ specific for each nanodiamond is obtained from the equivalent temperature at vanishing laser powers.

\section{Hydrodynamic Radius}
For a spherical particle of radius $r$ in a rarefied gas, the damping writes~\cite{Gieseler2012PRL}:
\begin{equation}
\Gamma = 0.619 \dfrac{9}{\sqrt{2\pi}\rho_\text{diam}}\sqrt{\dfrac{M}{N_A k_B T_0}}\dfrac{p_\text{gas}}{r}\, ,
\end{equation}
where $\rho_\text{diam}\approx 3500$~Kg/m$^3$ is the diamond density, $M$ the molar mass of air, $T_0$ the room temperature and $p_\text{gas}$ the pressure inside the vacuum chamber.

Experimentally, we measure the particle damping at a given pressure (typically some tens of hPa) from the linewidth of the particle dynamics power spectral density.
From this damping, assuming the particle to be a sphere, we determine the effective hydrodynamic radius $r_\text{hydro}$ of the particle~\cite{Hoang2016PRL,Riviere2022AP}, which is used in the main text. 
While not giving a precise physical quantity, the hydrodynamic radius provides an idea of the typical particle size.

\newpage
\section{Absorption cross-section fit models}%
\begin{figure}[htb!]
        \centering
        \includegraphics[width=0.6\textwidth]{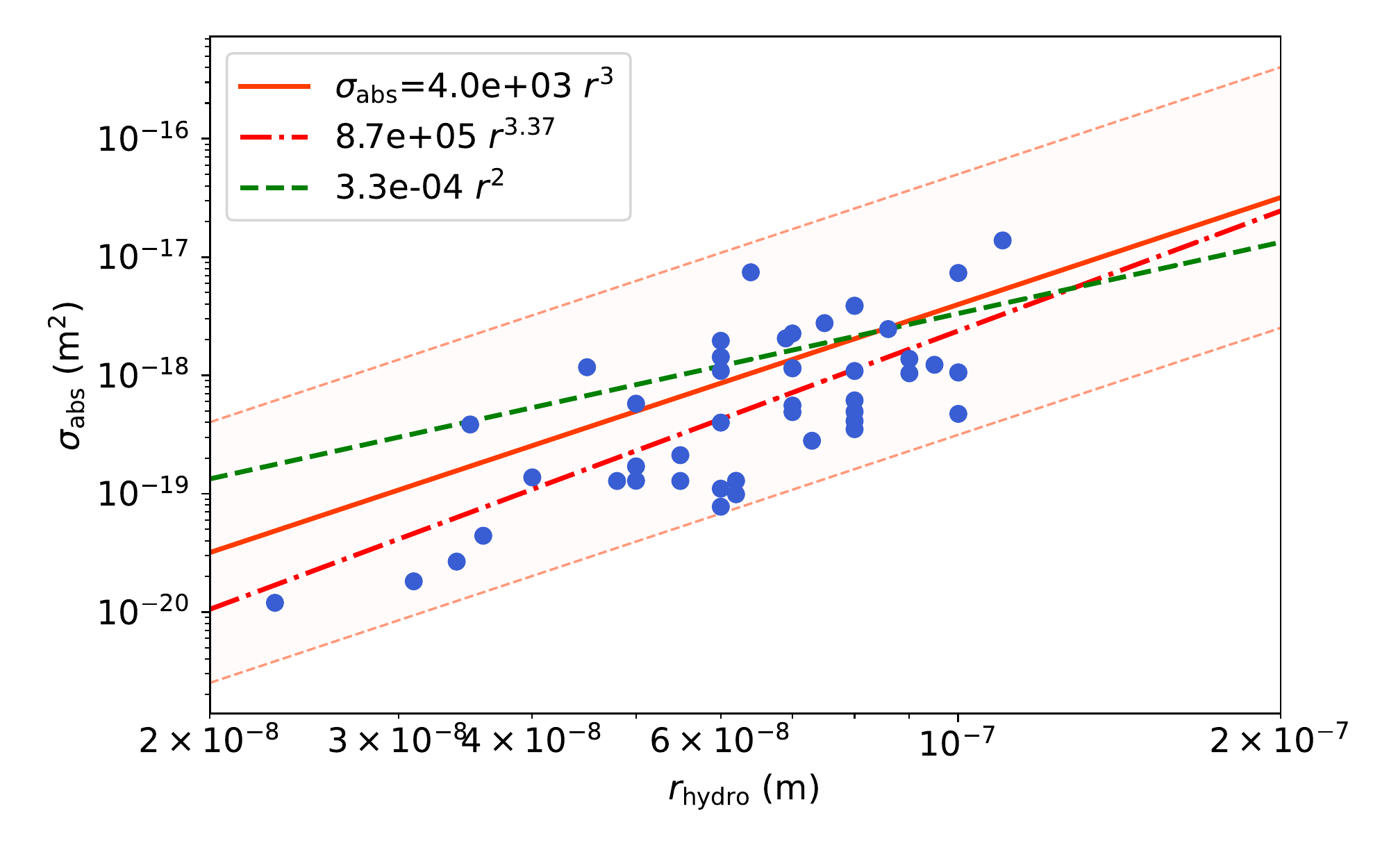}
        \caption{Comparison of fits of experimental data (blue dots) of absorption cross-section as a function of the particle hydrodynamic radius. The used fitting functions are $a_{h,n} \times r_\text{hydro}^n$ (red dotted lines), $a_{h,2} \times r_\text{hydro}^2$ (green dashed line), and $a_{h,3}\times r_\text{hydro}^3$ (orange line). The orange shadowed area corresponds to the $2\sigma$ error area compared to the $r^3$ fit, as in the main text.}
        \label{fig:SI_BestFit}
\end{figure}

\section{Sensitivity of diamond thermometry}
Ultimately the temperature sensitivity measured using ESR is given by~\cite{Dreau2011,Barry2020RMP}
\begin{equation}
        \eta_T \simeq \dfrac{\Gamma}{\mathcal C\sqrt{R}|\dfrac{\mathrm d D}{\mathrm d T}|}\, ,
\end{equation}
where $\Gamma$ is the ESR linewidth, $\mathcal C$ its contrast, $R$ the photon counting rate. 

In the experiments of optical levitation, we generally work with a moderate green excitation to keep a decent contrast, such that $R=200$~kcounts/s, and with a contrast ranging from 5 to 10~\%. For our HPHT diamonds, we observe $\Gamma\approx 10$~MHz. Close from ambient temperature, we have $\dfrac{\mathrm d D}{\mathrm d T}=-74$~kHz/K. Thus the sensitivity of our setup is typically on the order of a few K/$\sqrt{\mathrm{Hz}}$. Given our acquisition time of typically $1.5$~s/frequency this leads to a resolution of $\sim 1$~K.

\bibliography{biblio}